\documentclass[letterpaper, 12pt]{amsart}

\usepackage{orcidlink,thumbpdf,lmodern}

\usepackage{amsfonts}
\usepackage{amssymb}
\usepackage{amsmath,tabu,amsthm}
\usepackage{natbib}
\usepackage{ae,fancyvrb}

\usepackage[margin=1in]{geometry}

\usepackage{fixltx2e} 
\usepackage{listings} 
\usepackage{comment}
\usepackage{multicol} 
\newcommand{\eps}{\varepsilon}

\newcommand{\p}{\mathfrak p}
\newcommand{\q}{\mathfrak q}
\renewcommand{\aa}{\mathfrak a}

\newtheorem{theorem}{Theorem}
\newtheorem{proposition}[theorem]{Proposition}

\theoremstyle{definition}

\newcommand{\pkg}[1]{{\fontseries{m}\fontseries{b}\selectfont #1}}
\newcommand\code{\bgroup\@makeother\_\@makeother\~\@makeother\$\@codex}
\def\@codex#1{{\normalfont\ttfamily\hyphenchar\font=-1 #1}\egroup}
\let\proglang=\textsf
\let\code=\texttt

\DefineVerbatimEnvironment{Code}{Verbatim}{}
\DefineVerbatimEnvironment{CodeInput}{Verbatim}{fontshape=sl}
\DefineVerbatimEnvironment{CodeOutput}{Verbatim}{}
\newenvironment{CodeChunk}{}{}
\setkeys{Gin}{width=0.8\textwidth}

\author{Anna Bykhovskaya}

\address[Anna Bykhovskaya]
{Department of Economics\\
Duke University\\
USA\\
anna.bykhovskaya@duke.edu
}

\author{Vadim Gorin}

\address[Vadim Gorin]{
Departments of Statistics and Mathematics\\
University of California at Berkeley\\
USA\\
vadicgor@gmail.com
}

\author{Eszter Kiss}

\address[Eszter Kiss]
{
Department of Economics\\
Duke University\\
USA\\
ekiss2803@gmail.com
}

\title[Testing Large VARs for the Presence of Cointegration]{Largevars: An R Package for Testing Large VARs for the Presence of Cointegration}

\thanks{Vadim Gorin was partially supported by NSF grant DMS - 2246449.}

\begin{document}

\sloppy

\maketitle

\begin{abstract}
Cointegration is a property of multivariate time series that determines whether its non-stationary, growing components have a stationary linear combination.
\pkg{Largevars}  \proglang{R} package conducts a cointegration test for high-dimensional vector autoregressions of
order $k$ based on the large $N,T$ asymptotics of \citet{bykhovskaya_gorin_1, bykhovskaya_gorin_k}. The implemented test is a modification of the Johansen likelihood ratio test. In the absence of cointegration the test converges to the partial sum of the Airy$_1$ point process, an object arising in random matrix theory.

The package and this article contain simulated quantiles of the first ten partial sums of the Airy$_1$ point process that are precise up to the first $3$ digits. We also include two examples using \pkg{Largevars}: an empirical example on S\&P$100$ stocks and a simulated VAR(2) example.
\end{abstract}



\section{Introduction}

Vector Autoregressions (VARs) are a fundamental tool in econometrics and time series analysis, providing a framework for modeling the dynamic interrelationships among multiple time series. However, as the number of variables in a VAR increases, the complexity of the model grows significantly, posing challenges for both estimation and inference. One critical aspect of analyzing VAR models is testing for the presence of cointegration, which can inform whether a set of non-stationary series share a long-run equilibrium relationship. That is, whether a set of non-stationary time series has a stationary linear combination, see, e.g., \citet{johansen_book}.

There are several ways to test for the presence of cointegration (see, e.g., \citet{maddala} for the detailed description of various methods). Yet traditional tests for the presence of cointegration (e.g., likelihood ratio of \citet{johansen1988,johansen1991}) are not suitable for analyzing large systems, as they tend to significantly over-reject the null hypothesis, see, for example, \citet{ho_sorensen1996}, \citet{gonzalo_pitarakis}. To address this issue, \citet{bykhovskaya_gorin_1, bykhovskaya_gorin_k} propose an approach based on alternative asymptotics, where both the number of coordinates $N$ and the length $T$ of time series are large, and tailored for high-dimensional time series. \pkg{Largevars} package implements this approach.

The test implemented in the \pkg{Largevars} package is based on the squared sample canonical correlations between transformed past levels (lags) and changes (first differences) of the data, as outlined in Section \ref{Section_our_procedure}. Its asymptotic distribution (derived under $N,T\to\infty$ jointly and proportionally) is given by the partial sums of the Airy$_1$ point process, a random matrix object defined in Section \ref{section_tables}. The quantiles of the sums are necessary in order to implement the test; we have tabulated them and included both in the package and in the tables presented in this article. These tables are of independent interest and, with the exception of Table \ref{airy_quantiles1}, have not appeared in the literature before. Table \ref{airy_quantiles1} corresponds to the quantiles of the Tracy-Widom distribution, which were also tabulated in \citet{Bejan}.

Another \proglang{R} package that provides quantiles of random matrix origin is \pkg{RMTstat} \citep{rmstatpackage}, which offers density, distribution, and quantile functions for the Tracy–Widom distribution with parameters $\beta=1,2,4$, based on precomputed tables. In comparison, our package provides Tracy-Widom quantiles for $\beta=1$ in Table \ref{airy_quantiles1}, along with nine additional tables containing quantiles necessary for cointegration testing based on the Airy$_1$ partial sums. The method used in \pkg{RMTstat} to generate quantile tables is not applicable in our more general setting. Our alternative nontrivial algorithm, along with its \proglang{MATLAB} implementation, is described in detail in Section \ref{section_tables}.


Various \proglang{R} packages have been developed for cointegration testing. For instance, the \pkg{urca} package \citet{urcapackage} includes functions such as \code{ca.jo} (implementing the procedure of \citet{johansen1988, johansen1991}), \code{ca.po} (for the test of \citet{phillips1990asymptotic}), and \code{cajolst} (implementing the procedure of \citet{lutkepohl2004testing}). The \pkg{ARDL} package \citet{ardlpackage} provides functions \code{bounds\_f\_test} and \code{bounds\_t\_test}, which implement the Wald bounds-test and t-bounds test for no cointegration by \citet{pesaran2001bounds}. Additionally, the \pkg{bootCT} package \citet{bootctpackage} offers  \code{boot\_ardl} function for the bootstrap version of ARDL cointegration tests, as in \citet{bertelli2022bootstrap}. Several other implementations of cointegration tests exist in other software languages. Despite these offerings, the majority of existing packages are not tailored for high-dimensional settings, and their performance on time-series with large $N$ remains uncertain (see \citet{onatski_ecta, onatski_joe} for detailed theoretical discussions on over-rejection in classical tests for large $N$). Some of the packages explicitly prohibit the use of large $N$, e.g.\  \code{ca.jo} in the \pkg{urca} package does not output test results for $N>10$. Therefore, our new package complements existing software by providing specialized tools with theoretical assurances for high-dimensional settings.



\section{Cointegration test}
\label{Section_cointegration_test}

This section explains the theoretical foundations of the cointegration testing and challenges in their practical implementation.

\subsection{Setup and likelihood ratio test}

We consider a $N \times (T+1)$ data set represented by columns $X_t$, $0 \leq t \leq T$. These columns are interpreted as observations of an $N$-dimensional vector at $T+1$ time points or as $N$ scalar time series. $X_t$ is said to be \emph{cointegrated} if there exists a linear combination with coefficients $\beta$ of these time series such that the scalar time series $\beta^\top X_t$, for $0 \leq t \leq T$, is \emph{stationary} over time, potentially after detrending. Conversely, if every linear combination is non-stationary, we conclude no cointegration exists. In stationary scenarios, there is no growth over time, and correlations exhibit short-range behavior. Non-stationary scenarios, however, show growth and wider correlations, see \citet{brockwell1991time,johansen_book} for rigorous treatments. In particular, when $N=2$, cointegration implies a long-term equilibrium where the first and second coordinates of $X_t$ move together.

Cointegration has been extensively studied in econometrics, beginning with seminal works by \citet{granger1981,engle_granger1987}. Many variables in macroeconomics and finance, such as price levels, consumption, output, trade flows, and interest rates, may exhibit cointegration. A classic example is the relationship between interest rates for 3-month and 1-year US Treasury bills, which are cointegrated (see, e.g., \citet[Section 5.2]{BG_review} for an illustration). In portfolio management cointegrated stocks give rise to the strategy known as ``pairs trading''. This article and its accompanying software discuss statistical procedures for determining whether a given data set $X_t$ demonstrates cointegration. The main advantage of our approach is its applicability to settings with large $N$, in contrast to many existing packages and articles.

For the mathematical setup we assume that the data set is a realization of an $N$-dimensional vector autoregressive process of order $k$, denoted VAR($k$). The process is driven by a sequence of i.i.d.~mean-zero errors ${\eps_t}$ with non-degenerate covariance matrix $\Lambda$. In its error correction form, the model reads:
\begin{equation}\label{var_k}
\Delta X_t = \mu + \sum_{i=1}^{k-1} \Gamma_i \Delta X_{t-i} + \Pi X_{t-k} + \eps_t, \qquad t = 1, \ldots, T,
\end{equation}
where $\Delta X_t := X_t - X_{t-1}$. The parameters $\mu \in \mathbb{R}^N$, and $\Gamma_1, \ldots, \Gamma_{k-1}, \Pi \in \mathbb{R}^{N \times N}$ are unknown. The process is initialized with fixed values $X_{1-k}, \ldots, X_0$.

(Some authors use a slightly different form: $\Delta X_t=\mu+\Pi X_{t-1}+\sum\limits_{i=1}^{k-1}\tilde{\Gamma}_i\Delta X_{t-i}+\eps_t$, but the distinction is not crucial for our discussion.)

It is well known (see \citet{engle_granger1987,johansen_book}) that, under technical conditions, the process $X_t$ is cointegrated if and only if $\Pi\ne 0$. In particular, testing for the absence of cointegration can be recast as testing the hypothesis $\Pi = 0$ in model \eqref{var_k}.

A widely used procedure for testing this hypothesis is the Johansen likelihood ratio test, introduced in \citet{johansen1988,johansen1991} and based on Gaussian maximum likelihood (see also \citet{Anderson}). The method involves computing the eigenvalues $\lambda_1 \ge \lambda_2 \ge \dots \ge \lambda_N$ of a certain matrix constructed from the observed data ${X_t}$. These eigenvalues lie in the interval $[0,1]$ and can be interpreted as squared sample canonical correlations between transformed lagged levels and first differences of the time series. We describe a slightly modified version of this procedure in Section \ref{Section_our_procedure}.

The statistic for testing the hypothesis
$$H_0:\:{\rm rank}(\Pi)=0\qquad  (\text{i.e.},\,  \Pi\equiv0)$$
against the alternative
$$
 H(r):\quad {\rm rank}(\Pi)\in[1,r]
$$
is based on the $r$ largest eigenvalues and takes the form:
$$
 \sum_{i=1}^r\ln(1-\lambda_i).
$$
Under $H_0$ the $\lambda_i$'s tend to be small, making the statistic close to zero. Under the alternative $H(r)$, one expects some $\lambda_i$ to be close to one, which makes the sum more negative. Thus, the test rejects $H_0$ when the statistic is sufficiently negative.

The parameter $r$ is user-specified, and for fixed $N$ this gives rise to $N-1$ different tests corresponding to $r = 1, 2, \dots, N-1$. The asymptotic distribution of the statistic $\sum_{i=1}^r \ln(1 - \lambda_i)$ under $H_0$ as $T \to \infty$, with $N$ fixed was derived in \citet{johansen1988, johansen1991}. It involves the eigenvalues of a certain matrix of It\^{o} integrals. This result underpins the classical cointegration testing procedure: compute the test statistic from data and compare it to the quantiles of its theoretical asymptotic distribution under $H_0$; if the observed value is smaller than the critical threshold, reject $H_0$ and conclude that the series $X_t$ exhibits cointegration.

In practice the Johansen procedure and its associated software are typically applied only when $N$ is small. The main reason is that the quality of the approximation based on the asymptotic distribution involving It\^{o} integrals deteriorates rapidly as $N$ increases. Specifically, the distribution of the statistic $\sum_{i=1}^r \ln(1 - \lambda_i)$ can deviate significantly from its theoretical large-$T$ limit even for moderate values of $N$ (for example, the case $N = 10$, $T = 100$ already yields poor performance). As a result, the likelihood ratio test tends to over-reject the null hypothesis $H_0$ in finite samples. See \citet{onatski_ecta, onatski_joe} for a detailed theoretical analysis of this phenomenon.

This breakdown highlights the need to adapt the testing procedure and software for high-dimensional settings. In this work, we address this challenge by developing a new approach that is reliable for large $N$, building on recent theoretical results from \citet{bykhovskaya_gorin_1, bykhovskaya_gorin_k}.

\subsection{The procedure adapted for large $N$}
\label{Section_our_procedure}

We now discuss a procedure implemented in \pkg{Largevars} package, which is a modification of the construction of eigenvalues $\lambda_i$ used in the Johansen likelihood test. Starting with the data set $X_t$, $0\le t \le T$, we fix a number $r=1,2,\dots$ and perform the following steps.

\textbf{Step 1 (Detrending). } We de-trend and shift the data by defining
\begin{equation}
\label{eq_detrending}
 \tilde X_t = X_{t-1} - \frac{t-1}{T} (X_T-X_0), \qquad 1\le t \le T.
\end{equation}

\textbf{Step 2 (Cyclic indexing and regressor construction).} We define cyclic indices modulo $T$: for any $a\in\mathbb Z$, set
$$
 a\mid T= a+ m T,\quad \text{where } m\in\mathbb Z\text{ is such that } a+ m T\in \{1,2,\dots,T\}.
$$
Using this notation, we construct the following regressor matrices:
$$
 \tilde{Z}_{0t}=\Delta X_{t\mid T}\equiv\Delta X_{t},\quad\tilde{Z}_{kt}=\tilde X_{t-k+1\mid T},\quad\tilde{Z}_{1t}=(\Delta X_{t-1\mid T}^\top,\ldots,\Delta X_{t-k+1\mid T}^\top,1)^\top, \quad 1\le t \le T.
$$
Here and below $^\top$ denotes matrix transposition. For each fixed $t$,  the vector $\tilde{Z}_{1t}$  is a column of dimension $((k-1)N+1)\times 1$. The index $k$ in  $\tilde{Z}_{kt}$ is used symbolically to reflect the VAR($k$) structure and does not refer to a specific numerical value --- this convention follows \citet{johansen1988, johansen1991}.

We emphasize that due to the use of cyclic indices, values of $X_t$ at $t = 0, -1, \ldots$ are replaced by values at $t = T, T-1, \ldots$. However, when $k = 1$ (i.e., for a VAR($1$) model), no negative indices arise for $1 \le t \le T$, and the cyclic indexing becomes irrelevant.

\textbf{Step 3 (Regression residuals).} We compute the residuals from the regressions of $\tilde{Z}_{0t}$ and $\tilde{Z}_{kt}$ on $\tilde{Z}_{1t}$:
\begin{equation}\label{res_BG}
\tilde{R}_{it}=\tilde{Z}_{it}-\left(\sum\limits_{\tau=1}^{T} \tilde{Z}_{i\tau}\tilde{Z}_{1\tau}^\top\right)\left(\sum\limits_{\tau=1}^{T} \tilde{Z}_{1\tau}\tilde{Z}_{1\tau}^\top\right)^{-1}\tilde{Z}_{1t},\qquad i=0,k.
\end{equation}

\textbf{Step 4 (Canonical Correlations).} Let $\tilde{R}_i$ be the $N \times T$ matrix whose columns are $\tilde{R}_{it}$ for $1 \le t \le T$, for $i = 0, k$. Define the cross-product matrices
\begin{equation}\label{S_matrices}
\tilde{S}_{ij} = \sum_{t=1}^{T} \tilde{R}_{it} \tilde{R}_{jt}^\ast, \qquad i, j \in \{0, k\},
\end{equation}
and form the matrix
\begin{equation}\label{tilde_C}
\tilde{\mathcal{C}} = \tilde{S}_{k0} \tilde{S}_{00}^{-1} \tilde{S}_{0k} \tilde{S}_{kk}^{-1}.
\end{equation}
The eigenvalues $\tilde{\lambda}_1 \ge \ldots \ge \tilde{\lambda}_N$ of $\tilde{\mathcal{C}}$ represent the squared sample canonical correlations between $\tilde{R}_k$ and $\tilde{R}_0$. Equivalently, they solve the eigenvalue problem
\begin{equation}
\label{eq_vark_eig}
\det\left( \tilde{S}_{k0} \tilde{S}_{00}^{-1} \tilde{S}_{0k} - \tilde{\lambda} \tilde{S}_{kk} \right) = 0.
\end{equation}

\textbf{Step 5 (Test Statistic).} We construct the modified likelihood ratio statistic:
\begin{equation}\label{eq_LR_NT}
LR_{N,T}(r) = \sum_{i=1}^{r} \ln(1 - \tilde{\lambda}_i).
\end{equation}
The subscript $(N,T)$ indicates that this version of the Johansen LR test is tailored for the high-dimensional regime where both $N$ and $T$ are large. After centering and scaling, the statistic $LR_{N,T}(r)$ is compared with suitable critical values to decide whether to reject the null hypothesis $H_0$. Heuristically, rejection corresponds to the case where the largest eigenvalues $\tilde{\lambda}_i$ are significantly large and well-separated from the rest.

\bigskip

We now describe the asymptotic distribution theory underlying the critical values used in our procedure. These formulas are based on the limiting behavior of the test statistic $LR_{N,T}(r)$ as both $N, T \to \infty$. The relevant asymptotics were developed in \citet{bykhovskaya_gorin_1, bykhovskaya_gorin_k}, and we briefly recall the key results.

The limiting distribution involves a stochastic object known as the \emph{Airy$_1$ point process}, denoted by $\{\aa_i\}_{i=1}^{\infty}$. This is a random, strictly decreasing sequence of real numbers: $\aa_1 > \aa_2 > \aa_3 > \dots$. We discuss this process in more detail in the next section.

Let $T$, $N$, and $k$ be such that $\frac{T}{N} > k+1$. Define the following constants:
\begin{equation}
\label{eq_pq}
	\p=2, \qquad \q=\frac{T}{N}-k,\qquad \lambda_\pm=\frac{1}{(\p+\q)^2}\left[\sqrt{\p(\p+\q-1)}\pm \sqrt{\q}  \right]^2,
\end{equation}
$$
	c_1\left(N,T\right)=\ln\left(1-\lambda_+\right), \qquad
	c_2\left(N,T\right)=-\frac{2^{2/3} \lambda_+^{2/3}}{(1-\lambda_+)^{1/3} (\lambda_+-\lambda_-)^{1/3}} \left(\p+\q\right)^{-2/3}  <0.
$$
Then, under appropriate assumptions, it follows from \citet[Theorem 2]{bykhovskaya_gorin_1} and \citet[Theorem 9]{bykhovskaya_gorin_k} that
\begin{equation}
 \label{eq_asymptotic_approximation}
	 \frac{\sum_{i=1}^{r} \ln(1-\tilde{\lambda}_i)- r \cdot c_1(N,T)}{ N^{-2/3}  c_2(N,T)} \stackrel{d}{\longrightarrow} \sum_{i=1}^r \aa_i, \qquad N,T \to\infty.
\end{equation}
The practical applicability of this result depends on whether the theoretical assumptions hold for a given data set $X_t$. In Section \ref{Section_checks} we discuss model diagnostics that users can perform to assess the validity of the asymptotic approximation in real data.

To carry out the test in practice, one starts with the statistic $LR_{N,T}(r)$ defined in \eqref{eq_LR_NT}. We recommend choosing small values of $r$ (e.g., $r = 1, 2,$ or $3$). The approximation in \eqref{eq_asymptotic_approximation} assumes that $r$ is fixed as $N, T \to \infty$ --- the rationale for this and its implications are discussed in detail in \citet[Section 3.2]{bykhovskaya_gorin_1}.

The testing procedure then proceeds by computing the rescaled statistic:
$$
 \frac{LR_{N,T}(r)- r \cdot c_1(N,T)}{ N^{-2/3}  c_2(N,T)}
$$
and comparing it to the quantiles of the distribution $\sum_{i=1}^r \aa_i$. If the rescaled value exceeds the $\alpha$-quantile, we reject the null hypothesis of no cointegration at the $(1 - \alpha)$ significance level. The function \code{largevar()} in the \pkg{Largevars} package implements this procedure.

\subsection[Simulation of the Airy process]{Simulation of the Airy$_1$ point process}\label{section_tables}

The asymptotic formula \eqref{eq_asymptotic_approximation} shows that implementing our cointegration testing procedure requires knowledge of the distribution of the random variables $\sum_{i=1}^r \aa_i$. In this section we discuss how this distribution can be computed.

The Airy$_1$ point process is a random infinite sequence of real numbers $\aa_1 > \aa_2 > \aa_3 > \dots$ that can be defined via the following proposition:
\begin{proposition}[\citet{forrester1993}, \citet{tw1996}]  Let $Y_N$ be an $N\times N$ matrix of i.i.d.~$\mathcal{N}(0,2)$ Gaussian random variables, and let $\mu_{1;N}\ge \mu_{2;N}\ge \dots \mu_{N;N}$ be eigenvalues of $\frac{1}{2}\left(Y_N+Y_N^\top\right)$. Then, in the sense of convergence of finite-dimensional distributions,
	\begin{equation}
	\label{eq_GOE_to_Airy}
	\lim_{N\to\infty} \left\{N^{1/6}\left(\mu_{i;N}-2\sqrt{N}\right) \right\}_{i=1}^N = \{ \aa_i\}_{i=1}^\infty.
	\end{equation}
\end{proposition}

The distribution of $\aa_1$ is known as the Tracy–Widom distribution $F_1$. \citet{tw1996} showed that the cumulative distribution function of $F_1$ can be expressed as the solution to a Painlevé differential equation. Numerical solutions to this equation were used by \citet{Bejan} to compute highly accurate tables of quantiles of $F_1$. See also \citet{dieng2005distribution, bornemann2009numerical, trogdon2024computing} for further numerical advances.

Several software packages incorporate precomputed tables of the Tracy–Widom distribution for practical use. For example, the \pkg{RMTstat} package in \proglang{R} provides such functionality; see \citet{rmstatpackage}.

When $r > 1$, much less is known about the distribution of $\sum_{i=1}^r \aa_i$, and it is unclear whether any of the approaches described in the previous paragraph remain applicable. Therefore, in the \pkg{Largevars} package, we embed precomputed quantile tables for $\sum_{i=1}^r \aa_i$ obtained through direct simulation based on the definition in \eqref{eq_GOE_to_Airy}.

The convergence rate in \eqref{eq_GOE_to_Airy} is of order $N^{-1/3}$, which is relatively slow. For instance, even with a large matrix of size $N=1000$, the approximation error remains around $N^{-1/3} = 0.1$. \citet{johnstone2012fast} proposed computational techniques that accelerate convergence to $N^{-2/3}$ for $\aa_1$, but it is unknown whether such techniques yield similar improvements for $\sum_{i=1}^r \aa_i$ with $r > 1$. Moreover, testing this is nontrivial: while \citet{johnstone2012fast} could compare their numerics against the known distribution of $\aa_1$, no such benchmark exists for the sum of the top $r$ points. Consequently, to ensure accurate quantile estimation via \eqref{eq_GOE_to_Airy}, we opted for a very large matrix size: $N = 10^8$.

Using \eqref{eq_GOE_to_Airy} with $N = 10^8$ presents a computational challenge: no modern system can compute the eigenvalues of a $10^8 \times 10^8$ dense matrix. To circumvent this, we employ a numerical technique based on the tridiagonalization of the symmetric matrix $\frac{1}{2}(Y_N + Y_N^\top)$, following the approach of \citet{dumitriu_edelman}. This reduces the problem to computing the eigenvalues of a real symmetric tridiagonal matrix of size $N \times N$:
\begin{equation}
\label{eq_tridiag_matrix}
\begin{pmatrix}
\mathcal{N}(0,2) & \chi_{N-1} & 0 &  &  & 0\\
\chi_{N-1} & \mathcal{N}(0,2) & \chi_{N-2} &  &  & \\
0 & \chi_{N-2} & \mathcal{N}(0,2) & &  &  \\
 &  &  & \ddots &  & \\
 & & & & \mathcal{N}(0,2) & \chi_{1}\\
0 &  & & &\chi_{1} & \mathcal{N}(0,2)\\
\end{pmatrix},
\end{equation}
where all entries on or above the diagonal are independent. Here, $\mathcal{N}(0,2)$ denotes a normal random variable with mean $0$ and variance $2$, and $\chi_\ell$ denotes the square root of a chi-squared random variable with $\ell$ degrees of freedom.

Directly computing the eigenvalues of the full matrix in \eqref{eq_tridiag_matrix} for $N = 10^8$ remains computationally infeasible. However, one can instead consider the eigenvalues of its top-left $\sqrt{N} \times \sqrt{N}$ submatrix. Owing to the specific structure of \eqref{eq_tridiag_matrix}, the largest eigenvalues of the full $N \times N$ matrix and those of the $\sqrt{N} \times \sqrt{N}$ submatrix have the same asymptotic distribution; see \citet[Section 1.1]{edelman2005numerical} and \citet[Lemma 5.2]{johnstone2021spin} for theoretical justification.

We leverage this result in our simulations by performing $10^7$ Monte Carlo runs on symmetric tridiagonal random matrices of size $10^4 \times 10^4$, corresponding to the top-left corner of the tridiagonal matrix in \eqref{eq_tridiag_matrix} with $N = 10^8$. After computing their eigenvalues, we rescale them according to the transformation in \eqref{eq_GOE_to_Airy} with $N = 10^8$. This yields approximations for the distribution of the individual $\aa_i$ values and, consequently, for the sums $\sum_{i=1}^r \aa_i$. We carried out this procedure for $r = 1, 2, \dots, 10$.

To assess the quality of our approximation, we do not run all $10^7$ simulations in a single batch. Instead, we perform $10^6$ simulations ten times using different initial random seeds. The average of the resulting sample quantiles provides our estimates for the quantiles of $\sum_{i=1}^r \aa_i$, while the standard deviation across the ten runs (not shown here) offers a measure of their reliability. The resulting quantile tables are embedded within the package and also presented separately in Section \ref{Section_tables2}. The standard deviations suggest that the error is at most $\pm 1$ in the third significant digit, i.e., $\pm 0.01$ for $r=1$ and $\pm 0.1$ for $r=10$. For $r=1$, our results closely match those of \citet{Bejan}.

The simulation scripts were written in \proglang{MATLAB} and executed on a computing cluster provided by the Department of Economics at Duke University. Due to the substantial runtime, these simulations cannot be executed “on the fly” within the \proglang{R} package and also it limits our ability to further increase the precision of the quantile tables. For reproducibility, we include both the full script and a reduced version (with smaller $N$ and a fixed random seed), along with its output --- a low-precision version of the tables presented in Section \ref{Section_tables2}.

Finding higher-precision quantiles for $\sum_{i=1}^r \aa_i$ remains an open problem, whether by theoretical or numerical means. We hope that the tables in Section \ref{Section_tables2} will help stimulate further interest in this question.

\subsection{Model fit assessment}
\label{Section_checks}

The procedure for cointegration testing relies on the validity of the approximation \eqref{eq_asymptotic_approximation} under the null hypothesis $H_0$ of no cointegration. This approximation, in turn, depends on theoretical assumptions made in its derivation. A natural concern for users is whether these assumptions are reasonable for a given dataset $X_t$, $0 \le t \le T$.

\citet[Theorem 9]{bykhovskaya_gorin_k} provides a mathematical justification for \eqref{eq_asymptotic_approximation} under the setting where both $T$ and $N$ are large, with $T/N \in (k+1, \infty)$ bounded away from the endpoints. The proof assumes Gaussian errors $\varepsilon_t$ in \eqref{var_k} and imposes the restriction $\Gamma_1 = \Gamma_2 = \dots = \Gamma_{k-1} = 0$. The discussion following the theorem in the same article argues that the approximation should remain valid when the $\Gamma_i$ matrices are of low rank, and \citet[Section 7.1]{bykhovskaya_gorin_1} further argues that the Gaussianity assumption on $\varepsilon_t$ is likely not essential.

Moreover, \citet{onatski_ecta} and \citet[Theorem 3]{bykhovskaya_gorin_k} demonstrate that under these relaxed assumptions an additional result holds --- independent of whether cointegration is present (i.e., whether $\Pi = 0$ in \eqref{var_k} or not), as long as $\Pi$ remains of small rank. Specifically, they show that the histogram of \emph{all} eigenvalues $\tilde{\lambda}_1, \dots, \tilde{\lambda}_N$ from \eqref{eq_vark_eig} converges to a deterministic limiting distribution.

In more detail, the Wachter distribution is a probability distribution on the interval $[0,1]$ that depends on two parameters $\p > 1$ and $\q > 1$, and is defined by the density
\begin{equation}
\label{eq_Jacobi_equilibrium_1}
 \mu_{\p,\q}(x) = \frac{\p+\q}{2\pi} \cdot \frac{\sqrt{(x - \lambda_-)(\lambda_+ - x)}}{x(1 - x)} \, \mathbf{1}_{[\lambda_-, \lambda_+]}\,,
\end{equation}
where the support $[\lambda_-, \lambda_+] \subset (0,1)$ is given by
\begin{equation}
 \lambda_\pm = \frac{1}{(\p+\q)^2} \left( \sqrt{\p(\p+\q-1)} \pm \sqrt{\q} \right)^2.
\end{equation}

\citet{onatski_ecta} and \citet{bykhovskaya_gorin_k} show that if the parameters are chosen according to \eqref{eq_pq}, then the empirical distribution of the eigenvalues $\tilde{\lambda}_i$ from \eqref{eq_vark_eig} converges to the Wachter distribution:
\begin{equation}
\label{eq_Wachter_approx}
 \frac{1}{N} \sum_{i=1}^N \delta_{\tilde{\lambda}_i} \longrightarrow \mu_{\p,\q}(x)\, \mathrm{d}x, \qquad N, T \to \infty, \quad \text{weakly, in probability}.
\end{equation}

This convergence provides a practical check for the applicability of our procedure. If, for a given data set $X_t$, the histogram of eigenvalues $\tilde{\lambda}_i$ resembles the shape of the Wachter distribution (possibly excluding a few outlying values, which may correspond to cointegration), then it is reasonable to trust the assumptions behind our cointegration test. If not, the modeling assumptions are likely violated, and the test results should not be used.

To facilitate this diagnostic step, the function \code{largevar()} in our package includes an option to plot the empirical histogram of eigenvalues along with the Wachter density.

\section{Commands}

\subsection{Getting started}

The latest version of the \pkg{Largevars} package can always be found on Github and installed using  the \verb|devtools| R package:
\begin{Code}
	library(devtools)
	install_github("eszter-kiss/Largevars")
\end{Code}

The latest stable version from CRAN can be installed by \code{install.packages("Largevars")}. Help for using the functions in the package can be called by running ?? \verb|function name| . The empirical example in Section \ref{sectionSP100example} of this paper can provide further guidance.

\subsection[Function largevar]{Function \code{largevar}}

\code{largevar()} is the main function in the package that implements the cointegration test for high-dimensional VARs.
\begin{Code}
 largevar(data, k = 1, r = 1,  fin_sample_corr = FALSE,  plot_output = TRUE,
   significance_level = 0.05)
\end{Code}

\begin{multicols}{2}
    \code{data}
    \vfill\null
    \columnbreak
    A numeric matrix where the columns contain individual time series that will be examined for the presence of cointegrating relationships. The rows are indexed by $t=0,1,\ldots,T$ and the columns by $i=1,\ldots,N$. In the notations of Section \ref{Section_cointegration_test}, this is $X_t$, $0\le t\le T$.
\end{multicols}

\begin{multicols}{2}
    \code{k}
    \vfill\null
    \columnbreak
    The number of lags that we wish to employ in the vector autoregression, as in \eqref{var_k}. The default value is \code{k = 1}.
\end{multicols}

\begin{multicols}{2}
    \code{r}
    \vfill\null
    \columnbreak
    The number of largest eigenvalues used in the test as in \eqref{eq_LR_NT}. The default value is \code{r = 1}.
\end{multicols}

\begin{multicols}{2}
    \code{fin\_sample\_corr}
    \vfill\null
    \columnbreak
    A boolean variable indicating whether we wish to employ finite sample correction on our test statistic, as suggested in \citet[Discussion after Theorem 2 and Section 5.1]{bykhovskaya_gorin_1}, \citet[Footnote 13]{bykhovskaya_gorin_k}. The default value is \code{fin\_sample\_corr = FALSE}.
\end{multicols}

\begin{multicols}{2}
    \code{plot\_output}
    \vfill\null
    \columnbreak
    A boolean variable indicating whether we wish to generate a plot of the empirical distribution of eigenvalues discussed in Section \ref{Section_checks}. The default value is \code{plot\_output = TRUE}.
\end{multicols}

\begin{multicols}{2}
    \code{significance\_level}
    \vfill\null
    \columnbreak
    Specify the significance level at which the decision about the $H_0$ should be made. This is denoted $(1-\alpha)$ in Section \ref{Section_our_procedure}. The default value is \code{significance\_level = 0.05}.
\end{multicols}

The function \code{largevar()} operates according to the steps laid in out in Section \ref{Section_our_procedure}. The test statistic is formed based on the $r$ largest eigenvalues. Any value of $r$ can be used to reject the hypothesis  $H_0$ of no cointegration, and the user can try different options. We recommend small values such as $r=1,2,3$, see \citet[Section 3.2]{bykhovskaya_gorin_1} for the detailed discussion.

\code{largevar()} returns a list object that contains the test statistic, a statistical table with a subset of theoretical quantiles ($q=0.90,\,0.95,\,0.97,\,0.99$) presented for $r=1$ to $r=10$, the decision about $H_{0}$ at the significance level specified by the user, and the p-value. These can be accessed by \code{list\$statistic} (numeric value) \code{list\$significance\_test\$significance\_table} (numeric matrix), \code{list\$significance\_test\$boolean\_decision} (numeric value of $0$ or $1$, where $1$ means ``reject''), and \code{list\$significance\_test\$p\_value}  (numeric value), respectively.

The simulations for the quantiles of the limiting distribution were conducted for $r=1$ to $r=10$ values. For this reason, $p$~values are accessible at inputs $r=1$ to $r=10$ only. For larger $r$ inputs, the function returns the test statistic but not the $p$~value and not the decision about $H_{0}$ at the significance level specified by the user.

\subsection[Function quantile tables]{Function \code{quantile\_tables}}

To access the test quantile tables  for the partial sums of the Airy$_1$ point process, discussed in Section \ref{section_tables} and in Section \ref{Section_tables2}, the user can call the \code{quantile\_tables()} function. Quantile tables are available for $r=1$ to $r=10$. The function returns a numeric matrix, where the $0.ab$ quantile corresponds to the row $0.a$ and the column $b$. For example:

\begin{CodeChunk}
\begin{CodeInput}
R> quantile_tables(r=1)
\end{CodeInput}
\begin{CodeOutput}
        0      1      2      3      4      5      6      7      8      9
0.0   -Inf   -3.90  -3.61  -3.43  -3.30  -3.18  -3.08  -3.00  -2.92  -2.85
0.1   -2.78  -2.72  -2.67  -2.61  -2.56  -2.51  -2.46  -2.41  -2.37  -2.33
0.2   -2.29  -2.24  -2.20  -2.17  -2.13  -2.09  -2.05  -2.02  -1.98  -1.95
0.3   -1.91  -1.88  -1.84  -1.81  -1.78  -1.74  -1.71  -1.68  -1.65  -1.62
0.4   -1.58  -1.55  -1.52  -1.49  -1.46  -1.43  -1.40  -1.36  -1.33  -1.30
0.5   -1.27  -1.24  -1.21  -1.17  -1.14  -1.11  -1.08  -1.05  -1.01  -0.98
0.6   -0.95  -0.91  -0.88  -0.85  -0.81  -0.78  -0.74  -0.71  -0.67  -0.63
0.7   -0.59  -0.56  -0.52  -0.48  -0.44  -0.39  -0.35  -0.31  -0.26  -0.22
0.8   -0.17  -0.12  -0.07  -0.01   0.04   0.10   0.16   0.23   0.30   0.37
0.9    0.45   0.53   0.63   0.73   0.85   0.98   1.14   1.33   1.60   2.02
\end{CodeOutput}
\end{CodeChunk}

\subsection[Function sim function]{Function \code{sim\_function}}

\code{sim\_function()} is an auxiliary function that allows the user to calculate an empirical $p$~value based on a simulation of the data generating process $\widehat H_{0}$ stated in  of  \citet[equation (10)]{bykhovskaya_gorin_k}; equivalently this is the model \eqref{var_k} based on mean $0$ Gaussian errors $\eps_t$ and $\Gamma_1=\dots=\Gamma_{k-1}=0$. This function should be used only for \textit{quick approximate assessments}, as precise computation of the distribution of the test statistic requires a very large number of simulations, as discussed in Section \ref{section_tables}.

\begin{Code}
sim_function(N, tau, stat_value, k = 1,
  r = 1, fin_sample_corr = FALSE, sim_num = 1000,  seed)
\end{Code}

\begin{multicols}{2}
    \code{N}
    \vfill\null
    \columnbreak
    The number of time series used in simulations.
\end{multicols}

\begin{multicols}{2}
    \code{tau}
    \vfill\null
    \columnbreak
    The length of the time series used in simulations. If time is indexed as $t=0,1,\ldots,T$, then $\tau=T+1$.
\end{multicols}

\begin{multicols}{2}
    \code{stat\_value}
    \vfill\null
    \columnbreak
    The test statistic value for which the $p$~value is calculated.
\end{multicols}

\begin{multicols}{2}
    \code{k}
    \vfill\null
    \columnbreak
    The number of lags that we wish to employ in the vector autoregression. The default value is $k=1$.
\end{multicols}

\begin{multicols}{2}
    \code{r}
    \vfill\null
    \columnbreak
    The number of largest eigenvalues used in the test. The default value is $r=1$.
\end{multicols}

\begin{multicols}{2}
    \code{fin\_sample\_corr}
    \vfill\null
    \columnbreak
    A boolean variable indicating whether we wish to employ finite sample correction on our test statistics. The default value is \verb|fin_sample_corr=FALSE|.
\end{multicols}

\begin{multicols}{2}
    \code{sim\_num}
    \vfill\null
    \columnbreak
    The number of simulations that the function conducts for $H_{0}$. The default value is \verb|sim_num = 1000|.
\end{multicols}

\begin{multicols}{2}
    \code{seed}
    \vfill\null
    \columnbreak
    A numeric variable for the user to set to make simulations replicable. If not set by the user, there is no seed set for the simulations.
\end{multicols}

The function \code{sim\_function()} runs the cointegration test (following the steps of Section \ref{Section_cointegration_test}) on simulated data generated under the evolution \eqref{var_k}  based on mean $0$ Gaussian errors $\eps_t$ and $\Gamma_1=\dots=\Gamma_{k-1}=0$ and calculates the empirical $p$~value  based on the test statistic \eqref{eq_LR_NT} corresponding to \code{r} specified by the user. The empirical $p$~value is defined as the fraction of realizations larger than the specified \code{stat\_value}. For comparison purposes, it is advised to specify the same parameters \code{k} and \code{r} as one expects to use for the run of \code{largevar()} for the desired data set.

\code{sim\_function()} returns a list object that contains the simulation values, the empirical $p$~value and a histogram of the distribution of simulated test statistic values (which is an approximation of the probability distribution of the test statistic).

\section{Examples}

This section provides two examples of the usage of the package. Section \ref{sectionSP100example} replicates the S$\&$P$100$ example from \citet{bykhovskaya_gorin_1, bykhovskaya_gorin_k}, while Section \ref{section_simulated_example} uses simulated data. Both examples include the code, which can be copied into \proglang{R}.

\subsection[SP 100]{S\&P100}\label{sectionSP100example}
We use logarithms of weekly adjusted closing prices of assets in the S\&P100 over ten years (01.01.2010--01.01.2020), which gives us $\tau=522$ observations across time. The S\&P100 includes 101 stocks, with Google having two classes of stocks. We use 92 of those stocks, those for which data were available for our chosen time period. Only one of Google’s two listed stocks is kept in the sample. Therefore, $N = 92$, $T = 521$ and $T/N  \approx 5.66$. We obtained the raw data from Yahoo! Finance and made the sample available in the ``data'' folder of the package for convenient data loading:

\begin{Code}
data("s_p100_price")
\end{Code}

We first make necessary transformations, then convert to a numeric matrix to match function requirements:

\begin{Code}
dataSP <- log(s_p100_price[, seq(2, dim(s_p100_price)[2])])
dataSP <- as.matrix(dataSP)
\end{Code}

There is documentation available for the following function which can be called using

\begin{Code}
`?`(largevar)
\end{Code}

The following code conducts the cointegration test and displays its results:

\begin{Code}
result <- largevar(data = dataSP,
                   k = 1,
                   r = 1,
                   fin_sample_corr = FALSE,
                   plot_output = TRUE,
                   significance_level = 0.05)

result	
\end{Code}

Since we set \code{plot\_output=TRUE}, we obtain a histogram of eigenvalues solving \eqref{eq_vark_eig}, as shown in Figure \ref{Figure_Wacter}. The resemblance of the histogram with the theoretical curve is very good and we expect that our cointegration test should be applicable to this data set.
The remaining output of \code{largevar()} is displayed in the console as:

\begin{figure}[t]
\centering
\includegraphics[scale=0.7]{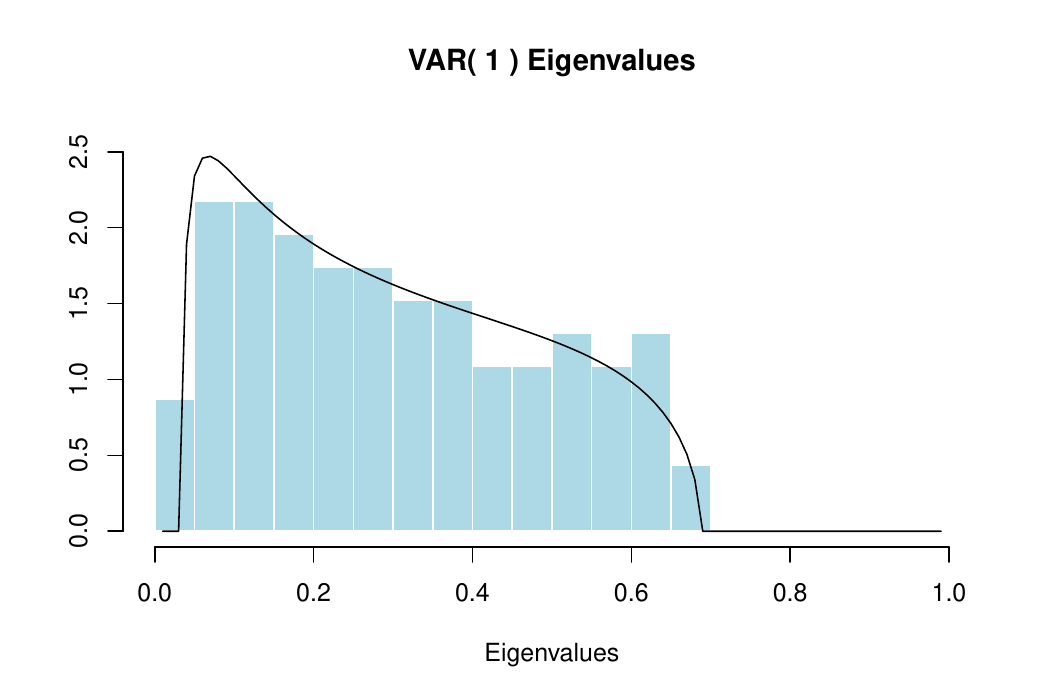}
\caption{\label{Figure_Wacter}Histogram of eigenvalues and Wachter distribution, as discussed in Section \ref{Section_checks}.}
\end{figure}

\begin{CodeOutput}
Output for the largevar function
=================================
Cointegration test for high-dimensional VAR(k)   T= 521 N= 92

 10% Crit. value 5% Crit. value 1% Crit. value Test stat.
            0.45           0.98           2.02      -0.28

If the test statistic is larger than the quantile, reject H0.
===============================================================
Test statistic: -0.2777314
The p-value is  0.23
Decision about H0:  0
\end{CodeOutput}

The decision $0$ means that we do not reject $H_0$, i.e., it is likely that the data set has no cointegration. If we want to individually access certain values from the output list, we can do it  by referencing the elements of the list:

\begin{CodeChunk}
\begin{CodeInput}
R> result$statistic
\end{CodeInput}
\begin{CodeOutput}
[1] -0.2777314
\end{CodeOutput}
\begin{CodeInput}
R> result$significance_test$p_value
\end{CodeInput}
\begin{CodeOutput}
[1] 0.23
\end{CodeOutput}
\begin{CodeInput}
R> result$significance_test$boolean_decision
\end{CodeInput}
\begin{CodeOutput}
[1] 0
\end{CodeOutput}
\begin{CodeInput}
R> result$significance_test$significance_table
\end{CodeInput}
\begin{CodeOutput}
        0.90    0.95    0.97    0.99   Test stat.
r=1     0.45    0.98    1.33    2.02   -0.2777314
r=2    -1.87   -1.09   -0.57    0.42   -1.4995879
r=3    -5.90   -4.90   -4.24   -2.99   -5.4154889
r=4   -11.35  -10.15   -9.37   -7.87  -10.5527603
r=5   -18.07  -16.69  -15.79  -14.07  -16.7460847
r=6   -25.95  -24.40  -23.38  -21.45  -23.2178976
r=7   -34.90  -33.19  -32.07  -29.95  -31.1080001
r=8   -44.88  -43.01  -41.79  -39.47  -39.3197363
r=9   -55.82  -53.80  -52.48  -49.99  -49.8419822
r=10  -67.70  -65.53  -64.12  -61.45  -60.4894485
\end{CodeOutput}
\end{CodeChunk}

We can further compare the exact $p$~value (which was outputted by \code{largevar()}) with a sample value obtained trough $1000$ simulations, by running \code{sim\_function()}:

\begin{Code}
result2 <- sim_function(N = 92,
                        tau = 522,
                        stat_value = result$statistic,
                        k = 1,
                        r = 1,
                        fin_sample_corr = FALSE,
                        sim_num = 1000,
                        seed = 333)
> result2

Output for the sim_function function
===================================
The empirical p-value is  0.247
\end{Code}

As we see, the empirical $p$~value $0.247$ is close to the previous output $0.23$, but there is a small mismatch, as expected.

\subsection{Simulation example}\label{section_simulated_example}

We also present an example based on simulated data that users can replicate. The code below generates VAR($2$) with $N = 100$, $T = 1500$, and
\begin{equation}\label{eq_simul_ex}
\begin{split}
   \begin{pmatrix} \Delta X_{1t}\\ \Delta X_{2t} \end{pmatrix}&= \begin{pmatrix} -0.9 & 0.8\\ 0& 0\end{pmatrix} \begin{pmatrix} X_{1t-2}\\ X_{2t-2} \end{pmatrix}+ \begin{pmatrix} -0.7& 0.8\\ 0& 0.3 \end{pmatrix} \begin{pmatrix}\Delta X_{1t-1}\\\Delta X_{2t-1} \end{pmatrix} +\begin{pmatrix} \eps_{1t}\\ \eps_{2t} \end{pmatrix},\,t=1,\ldots,T, \\
   \begin{pmatrix}\Delta X_{4t}\\\Delta X_{5t} \end{pmatrix}&= \begin{pmatrix}-0.9 & 0.8\\ 0& 0\end{pmatrix}\begin{pmatrix} X_{4t-2}\\ X_{5t-2} \end{pmatrix}+ \begin{pmatrix} -1.2& 0.8\\ 0& 0.25 \end{pmatrix} \begin{pmatrix}\Delta X_{4t-1}\\\Delta X_{5t-1} \end{pmatrix} +\begin{pmatrix} \eps_{4t}\\ \eps_{5t} \end{pmatrix},\,t=1,\ldots,T,\\
  \Delta X_{it}&=\eps_{it},\,   i\neq1,2,4,5, \,t=1,\ldots,T,
\end{split}
\end{equation}
where $\Delta X_{it}:=X_{it}- X_{it-1}$. The process is initialized by vectors $X_0,\,X_{-1}$ with independent standard normal coordinates. The data generating process \eqref{eq_simul_ex} corresponds to a matrix $\Pi$ of rank $2$: $\Pi$ has two nonzero and linearly independent rows. To be more precise, the coefficient matrices in Eq.~\eqref{eq_simul_ex} correspond to $N-2$ unit root and $2$ stationary components. We create a data set based on evolution \eqref{eq_simul_ex} and Gaussian errors, and then run the cointegration test on it, as follows.

The following code constructs matrices $\Pi$ and $\Gamma$ that by construction create two separate cointegrating systems, the first and second, and the fourth and fifth time series:
\begin{Code}
set.seed(333)
T_ <- 1500
N <- 100

Pi <- matrix(0, N, N)
Pi[1:5, 1:5] <- matrix(c(-0.9, rep(0, 4), 0.8, rep(0, 12), -0.9,
                         rep(0, 4), 0.8, 0), 5, 5)
Gamma <- matrix(0, N, N)
Gamma[1:5, 1:5] <- matrix(c(-0.7, rep(0, 4), 0.8, 0.3, rep(0, 11), -1.2,
                            rep(0, 4), 0.8, 0.25), 5, 5)
\end{Code}

The initialization of the time series by setting all the below values:
\begin{Code}
Xminus1 <- matrix(rnorm(N), N, 1)
X0 <- matrix(rnorm(N), N, 1)
dX <- matrix(0, N, T_)
dX0 <- X0 - Xminus1

epsilon <- matrix(rnorm(N * T_), N, T_)
dX[ , 1] <- Pi %*% Xminus1 + Gamma %*% dX0 + epsilon[ , 1]
dX[ , 2] <- Pi %*% X0 + Gamma %*% dX[ , 1] + epsilon[ , 2]
dX[ , 3] <- Pi %*% (X0 + dX[ , 1]) + Gamma %*% dX[ , 2] + epsilon[ , 3]


\end{Code}

The development of the system up to $T$ is calculated, starting with changes $dX$ for each $t$:
\begin{Code}
for (t in 4:T_) {
  dX[ , t] <- Pi %*% (X0 + rowSums(dX[ , 1:(t - 2)])) +
             Gamma %*% dX[ , t - 1] + epsilon[ , t]
}

data_sim <- matrix(0, N, T_ + 1)
data_sim[ , 1] <- X0
for (t in 2:(T_ + 1)) {
  data_sim[ , t] <- data_sim[ , t - 1] + dX[ , t - 1]
}
data_sim <- t(data_sim)
\end{Code}

Finally, we conduct the cointegration test and display the results:

\begin{Code}
result <- largevar(data = data_sim, k = 2, r = 2, fin_sample_corr = FALSE,
			plot_output = TRUE, significance_level = 0.05)
> result

Output for the largevar function
=================================
Cointegration test for high-dimensional VAR(k)   T= 1500 N= 100

 10% Crit. value 5% Crit. value 1% Crit. value Test stat.
           -1.87          -1.09           0.42      48.43

If the test statistic is larger than the quantile, reject H0.
===============================================================
Test statistic: 48.42677
The p-value is  0.01
Decision about H0:  1
\end{Code}

\begin{figure}[t]
\centering
\includegraphics[scale=0.7]{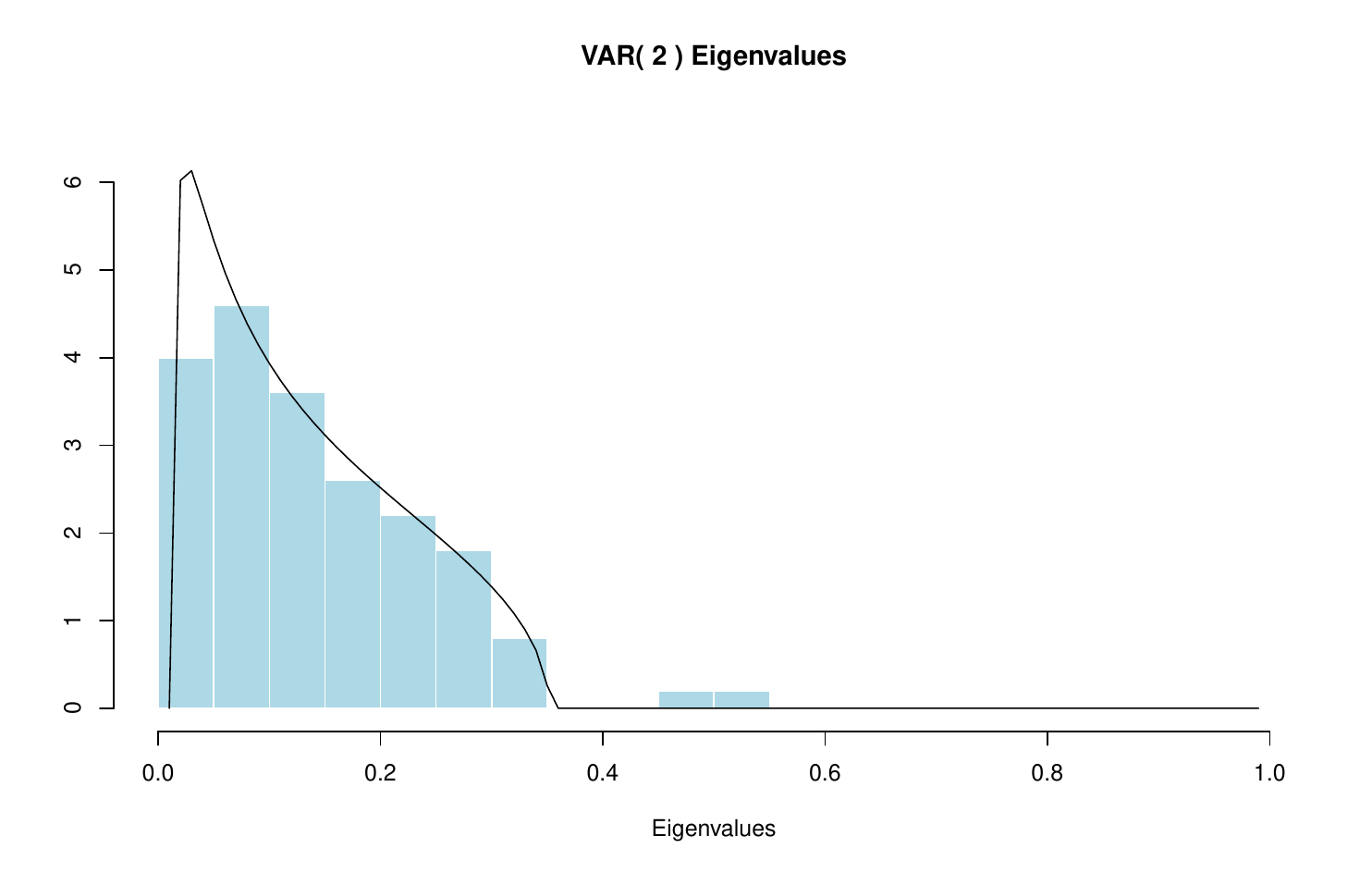}
\caption{\label{Figure_Wacter_2}Histogram of eigenvalues and Wachter distribution, as discussed in Section \ref{Section_checks}.}
\end{figure}

Since the decision is $1$, we reject $H_0$ and conclude that the data set has cointegration. The rejection is in line with the largest eigenvalue being significantly to the right from $\lambda_+$. The separation of eigenvalues is clearly visible in the histogram output, see Figure \ref{Figure_Wacter_2}.

If we want to take a look at how the significance of our test statistics vary across different choices of $r$, we can call the table below. The $p$~values for our test statistics stay below $0.01$.

\begin{CodeChunk}
\begin{CodeInput}
R> result$significance_test$significance_table
\end{CodeInput}
\begin{CodeOutput}
       0.90     0.95     0.99   Test stat.
r=1    0.45     0.98     2.02   27.357695
r=2   -1.87    -1.09     0.42   48.426766
r=3   -5.90    -4.90    -2.99   46.505972
r=4  -11.35   -10.15    -7.87   44.057939
r=5  -18.07   -16.69   -14.07   39.016668
r=6  -25.95   -24.40   -21.45   31.463442
r=7  -34.90   -33.19   -29.95   22.644198
r=8  -44.88   -43.01   -39.47   12.781779
r=9  -55.82   -53.80   -49.99    2.638057
r=10 -67.70   -65.53   -61.45   -7.878603
\end{CodeOutput}
\end{CodeChunk}

\section{Tables of quantiles}
\label{Section_tables2}

We include the tables discussed in Section \ref{section_tables}, both inside the package and in this section. These tables are used inside  \code{largevar()} to obtain the quantiles. The tables below present our simulation results. The $0.ab$ quantile in each table corresponds to the row $0.a$ and the column $b$.
The standard deviations of our results suggest that the error is at most $\pm 1$ in the third digit of the elements of the Airy$_1$ sequence, meaning that the error is $\pm 0.01$ for $r=1$ and $\pm 0.1$ for $r=10$.

\bigskip

\begin{table}[h!]
	\begin{tabular}{c|c|c|c|c|c|c|c|c|c|c|}
		\hline
		q & $0$ & $1$ & $2$ & $3$ & $4$ & $5$ & $6$ & $7$ & $8$ & $9$\\
		\hline
	    0.0 & $- \infty$   & -3.90 & -3.61 & -3.43 & -3.30 & -3.18 & -3.08 & -3.00 & -2.92 & -2.85\\
		\hline
        0.1 & -2.78 & -2.72 & -2.67 & -2.61 & -2.56 & -2.51 & -2.46 & -2.41 & -2.37 & -2.33\\
        \hline
        0.2 &  -2.29 & -2.24 & -2.20 & -2.17 & -2.13 & -2.09 & -2.05 & -2.02 & -1.98 & -1.95\\
        \hline
        0.3 &  -1.91 & -1.88 & -1.84 & -1.81 & -1.78 & -1.74 & -1.71 & -1.68 & -1.65 & -1.62\\
		\hline
        0.4 &  -1.58 & -1.55 & -1.52 & -1.49 & -1.46 & -1.43 & -1.40 & -1.36 & -1.33 & -1.30\\
        \hline
        0.5 & -1.27 & -1.24 & -1.21 & -1.17 & -1.14 & -1.11 & -1.08 & -1.05 & -1.01 & -0.98\\
        \hline
        0.6 & -0.95 & -0.91 & -0.88 & -0.85 & -0.81 & -0.78 & -0.74 & -0.71 & -0.67 & -0.63\\
		\hline
        0.7 &  -0.59 & -0.56 & -0.52 & -0.48 & -0.44 & -0.39 & -0.35 & -0.31 & -0.26 & -0.22\\
        \hline
        0.8 &  -0.17 & -0.12 & -0.07 & -0.01 & 0.04 & 0.10 & 0.16 & 0.23 & 0.30 & 0.37\\
        \hline
        0.9 &  0.45 & 0.53 & 0.63 & 0.73 & 0.85 & 0.98 & 1.14 & 1.33 & 1.60 & 2.02\\
        \hline
        \hline
	\end{tabular}
	\caption{Quantiles of $\aa_1$ based on $10^7$ Monte Carlo simulations of \mbox{$10^8\times 10^8$} tridiagonal matrices.}	
  \label{airy_quantiles1}
\end{table}

\begin{table}[h!]
	\begin{tabular}{c|c|c|c|c|c|c|c|c|c|c|}
		\hline
		q & $0$ & $1$ & $2$ & $3$ & $4$ & $5$ & $6$ & $7$ & $8$ & $9$\\
		\hline
	   0.0 &  $- \infty$ & -8.93 & -8.44 & -8.12 & -7.88 & -7.69 & -7.52 & -7.37 & -7.24 & -7.12 \\
		\hline
0.1 & -7.01 & -6.91 & -6.81 & -6.72 & -6.63 & -6.54 & -6.46 & -6.39 & -6.31 & -6.24\\
		\hline
0.2 & -6.17 & -6.10 & -6.04 & -5.97 & -5.91 & -5.85 & -5.79 & -5.73 & -5.67 & -5.61\\
		\hline
0.3 & -5.56 & -5.50 & -5.45 & -5.39 & -5.34 & -5.29 & -5.23 & -5.18 & -5.13 & -5.08\\
		\hline
0.4 & -5.03 & -4.97 & -4.92 & -4.87 & -4.82 & -4.77 & -4.72 & -4.67 & -4.62 & -4.57\\
		\hline
0.5 & -4.52 & -4.47 & -4.42 & -4.37 & -4.32 & -4.27 & -4.22 & -4.17 & -4.11 & -4.06\\
		\hline
0.6 & -4.01 & -3.96 & -3.91 & -3.85 & -3.80 & -3.74 & -3.69 & -3.63 & -3.57 & -3.52\\
		\hline
0.7 & -3.46 & -3.40 & -3.34 & -3.27 & -3.21 & -3.15 & -3.08 & -3.01 & -2.94 & -2.87\\
		\hline
0.8 & -2.80 & -2.72 & -2.65 & -2.57 & -2.48 & -2.39 & -2.30 & -2.20 & -2.10 & -1.99\\
		\hline
0.9 & -1.87 & -1.75 & -1.61 & -1.46 & -1.29 & -1.09 & -0.86 & -0.57 & -0.19 & 0.42\\
		\hline
        \hline
	\end{tabular}
	\caption{Quantiles of $\aa_1+\aa_2$ based on $10^7$ Monte Carlo simulations of \mbox{$10^8\times 10^8$}  tridiagonal matrices.}	
  \label{airy_quantiles2}
\end{table}

\begin{table}[h!]
	\begin{tabular}{c|c|c|c|c|c|c|c|c|c|c|}
		\hline
		q & $0$ & $1$ & $2$ & $3$ & $4$ & $5$ & $6$ & $7$ & $8$ & $9$\\
		\hline
	   0.0 &  $- \infty$ & -15.2 & -14.6 & -14.1 & -13.8 & -13.5 & -13.3 & -13.1 & -12.9 & -12.8\\
	\hline

0.1 & -12.6 & -12.5 & -12.4 & -12.2 & -12.1 & -12.0 & -11.9 & -11.8 & -11.7 & -11.6\\
	\hline

0.2 & -11.5 & -11.4 & -11.3 & -11.3 & -11.2 & -11.1 & -11.0 & -10.9 & -10.9 & -10.8\\
	\hline

0.3 & -10.7 & -10.6 & -10.6 & -10.5 & -10.4 & -10.3 & -10.3 & -10.2 & -10.1 & -10.1\\
	\hline

0.4 & -10.0 & -9.93 & -9.87 & -9.80 & -9.73 & -9.67 & -9.60 & -9.54 & -9.47 & -9.40\\
	\hline

0.5 & -9.34 & -9.27 & -9.21 & -9.14 & -9.07 & -9.01 & -8.94 & -8.87 & -8.80 & -8.74\\
	\hline

0.6 & -8.67 & -8.60 & -8.53 & -8.46 & -8.39 & -8.32 & -8.25 & -8.17 & -8.10 & -8.02\\
	\hline

0.7 & -7.95 & -7.87 & -7.79 & -7.71 & -7.63 & -7.55 & -7.46 & -7.37 & -7.28 & -7.19\\
	\hline

0.8 & -7.10 & -7.00 & -6.90 & -6.79 & -6.68 & -6.57 & -6.45 & -6.33 & -6.19 & -6.05\\
	\hline

0.9 & -5.90 & -5.74 & -5.56 & -5.37 & -5.15 & -4.90 & -4.60 & -4.24 & -3.76 & -2.99\\
	\hline
        \hline
	\end{tabular}
	\caption{Quantiles of $\sum_{i=1}^3 \aa_i$ based on $10^7$ Monte Carlo simulations of \mbox{$10^8\times 10^8$}  tridiagonal matrices.}	
  \label{airy_quantiles3}
\end{table}

\begin{table}[h!]
	\begin{tabular}{c|c|c|c|c|c|c|c|c|c|c|}
		\hline
		q & $0$ & $1$ & $2$ & $3$ & $4$ & $5$ & $6$ & $7$ & $8$ & $9$\\
		\hline
	   0.0 & $- \infty$  & -22.7 & -21.9 & -21.4 & -21.0 & -20.7 & -20.4 & -20.1 & -19.9 & -19.7\\
 \hline
0.1 & -19.5 & -19.3 & -19.2 & -19.0 & -18.9 & -18.8 & -18.6 & -18.5 & -18.4 & -18.3\\
 \hline
0.2 & -18.2 & -18.0 & -17.9 & -17.8 & -17.7 & -17.6 & -17.5 & -17.4 & -17.3 & -17.3\\
 \hline
0.3 & -17.2 & -17.1 & -17.0 & -16.9 & -16.8 & -16.7 & -16.6 & -16.6 & -16.5 & -16.4\\
\hline
0.4 & -16.3 & -16.2 & -16.1 & -16.1 & -16.0 & -15.9 & -15.8 & -15.7 & -15.7 & -15.6\\
\hline
0.5 & -15.5 & -15.4 & -15.3 & -15.3 & -15.2 & -15.1 & -15.0 & -14.9 & -14.8 & -14.8\\
 \hline
0.6 & -14.7 & -14.6 & -14.5 & -14.4 & -14.4 & -14.3 & -14.2 & -14.1 & -14.0 & -13.9\\
 \hline
0.7 & -13.8 & -13.7 & -13.6 & -13.5 & -13.4 & -13.3 & -13.2 & -13.1 & -13.0 & -12.9\\
 \hline
0.8 & -12.8 & -12.7 & -12.6 & -12.4 & -12.3 & -12.2 & -12.0 & -11.9 & -11.7 & -11.5\\
 \hline
0.9 & -11.4 & -11.2 & -11.0 & -10.7 & -10.5 & -10.2 & -9.80 & -9.37 & -8.79 & -7.87\\
 \hline
        \hline
 \end{tabular}
	\caption{Quantiles of $\sum_{i=1}^4 \aa_i$ based on $10^7$ Monte Carlo simulations of \mbox{$10^8\times 10^8$}  tridiagonal matrices.}	
  \label{airy_quantiles4}
\end{table}

\begin{table}[h!]
	\begin{tabular}{c|c|c|c|c|c|c|c|c|c|c|}
		\hline
		q & $0$ & $1$ & $2$ & $3$ & $4$ & $5$ & $6$ & $7$ & $8$ & $9$\\
		\hline
	   0.0 & $- \infty$  & -31.3 & -30.3 & -29.7 & -29.2 & -28.9 & -28.5 & -28.2 & -28.0 & -27.8\\
\hline
0.1 & -27.5 & -27.4 & -27.2 & -27.0 & -26.8 & -26.7 & -26.5 & -26.4 & -26.2 & -26.1\\
\hline
0.2 & -26.0 & -25.8 & -25.7 & -25.6 & -25.5 & -25.3 & -25.2 & -25.1 & -25.0 & -24.9\\
\hline
0.3 & -24.8 & -24.7 & -24.6 & -24.5 & -24.4 & -24.3 & -24.2 & -24.1 & -24.0 & -23.9\\
\hline
0.4 & -23.8 & -23.7 & -23.6 & -23.5 & -23.4 & -23.3 & -23.2 & -23.1 & -23.1 & -23.0\\
\hline
0.5 & -22.9 & -22.8 & -22.7 & -22.6 & -22.5 & -22.4 & -22.3 & -22.2 & -22.1 & -22.0\\
\hline
0.6 & -21.9 & -21.8 & -21.7 & -21.6 & -21.5 & -21.4& -21.3 & -21.2 & -21.1 & -21.0\\
\hline
0.7 & -20.9 & -20.8 & -20.7 & -20.6 & -20.5 & -20.4 & -20.2 & -20.1 & -20.0 & -19.9\\
\hline
0.8 & -19.7 & -19.6 & -19.5 & -19.3 & -19.2 & -19.0 & -18.8 & -18.7 & -18.5 & -18.3\\
\hline
0.9 & -18.1 & -17.9 & -17.6 & -17.3 & -17.0 & -16.7 & -16.3 & -15.8 & -15.1 & -14.1\\
	\hline
        \hline
       	\end{tabular}
	\caption{Quantiles of $\sum_{i=1}^5 \aa_i$ based on $10^7$ Monte Carlo simulations of \mbox{$10^8\times 10^8$}  tridiagonal matrices.}	
  \label{airy_quantiles5}
\end{table}

\begin{table}[h!]
	\begin{tabular}{c|c|c|c|c|c|c|c|c|c|c|}
		\hline
		q & $0$ & $1$ & $2$ & $3$ & $4$ & $5$ & $6$ & $7$ & $8$ & $9$\\
		\hline
	    0.0 &  $- \infty$ & -40.9 & -39.8 & -39.1 & -38.6 & -38.1 & -37.8 & -37.4 & -37.2 & -36.9\\
\hline
0.1 & -36.7 & -36.4 & -36.2 & -36.0 & -35.8 & -35.6 & -35.5 & -35.3 & -35.1 & -35.0\\
\hline
0.2 & -34.8 & -34.7 & -34.6 & -34.4 & -34.3 & -34.2 & -34.0 & -33.9 & -33.8 & -33.7\\
\hline
0.3 & -33.5 & -33.4 & -33.3 & -33.2 & -33.1 & -33.0 & -32.9 & -32.7 & -32.6 & -32.5\\
\hline
0.4 & -32.4 & -32.3 & -32.2 & -32.1 & -32.0 & -31.9 & -31.8 & -31.7 & -31.6 & -31.5\\
\hline
0.5 & -31.4 & -31.3 & -31.1 & -31.0 & -30.9 & -30.8 & -30.7 & -30.6 & -30.5 & -30.4\\
\hline
0.6 & -30.3 & -30.2 & -30.1 & -30.0 & -29.9 & -29.7 & -29.6 & -29.5 & -29.4 & -29.3\\
\hline
0.7 & -29.1 & -29.0 & -28.9 & -28.8 & -28.7 & -28.5 & -28.4 & -28.3 & -28.1 & -28.0\\
\hline
0.8 & -27.8 & -27.7 & -27.5 & -27.3 & -27.2 & -27.0 & -26.8 & -26.6 & -26.4 & -26.2\\
\hline
0.9 & -29.0 & -25.7 & -25.4 & -25.1 & -24.8 & -24.4 & -23.9 & -23.4 & -22.6 & -21.5\\
\hline
        \hline
	\end{tabular}
	\caption{Quantiles of $\sum_{i=1}^6 \aa_i$ based on $10^7$ Monte Carlo simulations of \mbox{$10^8\times 10^8$}  tridiagonal matrices.}	
  \label{airy_quantiles6}
\end{table}

\begin{table}[h!]
	\begin{tabular}{c|c|c|c|c|c|c|c|c|c|c|}
		\hline
		q & $0$ & $1$ & $2$ & $3$ & $4$ & $5$ & $6$ & $7$ & $8$ & $9$\\
		\hline
	    0.0 &  $- \infty$ & -51.5 & -50.3 & -49.5 & -48.9 & -48.4 & -48.0 & -47.6 & -47.3 & -47.0\\
\hline
0.1 & -46.8 & -46.5 & -46.3 & -46.1 & -45.8 & -45.6 & -45.5 & -45.3 & -45.1 & -44.9\\
\hline
0.2 & -44.8 & -44.6 & -44.4 & -44.3 & -44.1 & -44.0 & -43.9 & -43.7 & -43.6 & -43.4\\
\hline
0.3 & -43.3 & -43.2 & -43.0 & -42.9 & -42.8 & -42.7 & -42.5 & -42.4 & -42.3 & -42.2\\
\hline
0.4 & -42.1 & -41.9 & -41.8 & -41.7 & -41.6 & -41.5 & -41.4 & -41.2 & -41.1 & -41.0\\
\hline
0.5 & -40.9 & -40.8 & -40.7 & -40.5 & -40.4 & -40.3 & -40.2 & -40.1 & -40.0 & -39.8\\
\hline
0.6 & -39.7 & -39.6 & -39.5 & -39.4 & -39.2 & -39.1 & -39.0 & -38.8 & -38.7 & -38.6\\
\hline
0.7 & -38.5 & -38.3 & -38.2 & -38.0 & -37.9 & -37.8 & -37.6 & -37.5 & -37.3 & -37.3\\
\hline
0.8 & -37.0 & -36.8 & -36.6 & -36.4 & -36.3 & -36.1 & -35.9 & -35.6 & -35.4 & -35.2\\
\hline
0.9 & -34.9 & -34.6 & -34.3 & -34.0 & -33.6 & -33.2 & -32.7 & -32.1 & -31.2 & -30.0\\
	\hline
        \hline
	\end{tabular}
	\caption{Quantiles of $\sum_{i=1}^7 \aa_i$ based on $10^7$ Monte Carlo simulations of \mbox{$10^8\times 10^8$} tridiagonal matrices.\label{airy_quantiles7}}	
\end{table}

\begin{table}[h!]
	\begin{tabular}{c|c|c|c|c|c|c|c|c|c|c|}
		\hline
		q & $0$ & $1$ & $2$ & $3$ & $4$ & $5$ & $6$ & $7$ & $8$ & $9$\\
		\hline
	    0.0 & $- \infty$  & -63.0 & -61.7 & -60.8 & -60.2 & -59.7 & -59.2 & -58.8 & -58.5 & -58.1\\
\hline
0.1 & -57.8 & -57.6 & -57.3 & -57.1 & -56.8 & -56.6 & -56.4 & -56.2 & -56.0 & -55.8\\
\hline
0.2 & -55.6 & -55.5 & -55.3 & -55.1 & -55.0 & -54.8 & -54.7 & -54.5 & -54.4 & -54.2\\
\hline
0.3 & -54.1 & -53.9 & -53.8 & -53.6 & -53.5 & -53.4 & -53.2 & -53.1 & -53.0 & -52.8\\
\hline
0.4 & -52.7 & -52.6 & -52.4 & -52.3 & -52.2 & -52.0 & -51.9 & -51.8 & -51.7 & -51.5\\
\hline
0.5 & -51.4 & -51.3 & -51.2 & -51.0 & -50.9 & -50.8 & -50.6 & -50.5 & -50.4 & -50.3\\
\hline
0.6 & -50.1 & -50.0 & -49.9 & -49.7 & -49.6 & -49.5 & -49.3 & -49.2 & -49.0 & -48.9\\
\hline
0.7 & -48.8 & -48.6 & -48.5 & -48.3 & -48.1 & -48.0 & -47.8 & -47.7 & -47.5 & -47.3\\
\hline
0.8 & -47.1 & -47.0 & -46.8 & -46.6 & -46.4 & -46.1 & -45.9 & -45.7 & -45.4 & -45.2\\
\hline
0.9 & -44.9 & -44.6 & -44.2 & -43.9 & -43.5 & -43.0 & -42.5 & -41.8 & -40.9 & -39.5\\
\hline
        \hline
	\end{tabular}
	\caption{Quantiles of $\sum_{i=1}^8 \aa_i$ based on $10^7$ Monte Carlo simulations of \mbox{$10^8\times 10^8$}  tridiagonal matrices.}	
  \label{airy_quantiles8}
\end{table}

\begin{table}[h!]
	\begin{tabular}{c|c|c|c|c|c|c|c|c|c|c|}
		\hline
		q & $0$ & $1$ & $2$ & $3$ & $4$ & $5$ & $6$ & $7$ & $8$ & $9$\\
		\hline
	   0.0 & $- \infty$  & -75.5 & -74.0 & -73.1 & -72.4 & -71.8 & -71.3 & -70.9 & -70.5 & -70.2\\
\hline
0.1 & -69.9 & -69.6 & -69.3 & -69.0 & -68.8 & -68.5 & -68.3 & -68.1 & -67.9 & -67.7\\
\hline
0.2 & -67.5 & -67.3 & -67.1 & -66.9 & -66.7 & -66.6 & -66.4 & -66.2 & -66.1 & -65.9\\
\hline
0.3 & -65.7 & -65.6 & -65.4 & -65.3 & -65.1 & -65.0 & -64.8 & -64.7 & -64.6 & -64.4\\
\hline
0.4 & -64.3 & -64.1 & -64.0 & -63.8 & -63.7 & -63.6 & -63.4 & -63.3 & -63.2 & -63.0\\
\hline
0.5 & -62.9 & -62.7 & -62.6 & -62.5 & -62.3 & -62.2 & -62.1 & -61.9 & -61.8 & -61.6\\
\hline
0.6 & -61.5 & -61.4 & -61.2 & -61.1 & -60.9 & -60.8 & -60.6 & -60.5 & -60.3 & -60.2\\
\hline
0.7 & -60.0 & -59.9 & -59.7 & -59.5 & -59.4 & -59.2 & -59.0 & -58.8 & -58.6 & -58.5\\
\hline
0.8 & -58.3 & -58.1 & -57.9 & -57.6 & -57.4 & -57.2 & -56.9 & -56.7 & -56.4 & -56.1\\
\hline
0.9 & -55.8 & -55.5 & -55.1 & -54.7 & -54.3 & -53.8 & -53.2 & -52.5& -51.5 & -50.0\\
\hline
        \hline
	\end{tabular}
	\caption{Quantiles of $\sum_{i=1}^9 \aa_i$ based on $10^7$ Monte Carlo simulations of \mbox{$10^8\times 10^8$}  tridiagonal matrices.}	
  \label{airy_quantiles9}
\end{table}

\begin{table}[h!]
	\begin{tabular}{c|c|c|c|c|c|c|c|c|c|c|}
		\hline
		q & $0$ & $1$ & $2$ & $3$ & $4$ & $5$ & $6$ & $7$ & $8$ & $9$\\
		\hline
	    0.0 & $- \infty$  & -88.8 & -87.2 & -86.2 & -85.5 & -84.9 & -84.3 & -83.9 & -83.5 & -83.1\\
\hline
0.1 & -82.8 & -82.4 & -82.1 & -81.8 & -81.6 & -81.3 & -81.1 & -80.8 & -80.6 & -80.4\\
\hline
0.2 & -80.2 & -80.0 & -79.8 & -79.6 & -79.4 & -79.2 & -79.0 & -78.9 & -78.7 & -78.5\\
\hline
0.3 & -78.3 & -78.2 & -78.0 & -77.8 & -77.7 & -77.5 & -77.4 & -77.2 & -77.1 & -76.9\\
\hline
0.4 & -76.8 & -76.6 & -76.5 & -76.3 & -76.2 & -76.0 & -75.9 & -75.7 & -75.6 & -75.4\\
\hline
0.5 & -75.3 & -75.1 & -75.0 & -74.8 & -74.7 & -74.5 & -74.4 & -74.2 & -74.1 & -73.9\\
\hline
0.6 & -73.8 & -73.6 & -73.5 & -73.3 & -73.2 & -73.0 & -72.8 & -72.7 & -72.5 & -72.4\\
\hline
0.7 & -72.2 & -72.0 & -71.8 & -71.7 & -71.5 & -71.3 & -71.1 & -70.9 & -70.7 & -70.5\\
\hline
0.8 & -70.3 & -70.1 & -69.9 & -69.6 & -69.4 & -69.2 & -68.9 & -68.6 & -68.3 & -68.0\\
\hline
0.9 & -67.7 & -67.3 & -67.0 & -66.5 & -66.1 & -65.5 & -64.9 & -64.1 & -63.1 & -61.5\\
\hline
        \hline
	\end{tabular}
	\caption{Quantiles of $\sum_{i=1}^{10} \aa_i$based on $10^7$ Monte Carlo simulations of \mbox{$10^8\times 10^8$}  tridiagonal matrices.}	
  \label{airy_quantiles10}
\end{table}

\newpage

\bibliographystyle{abbrvnat}
\bibliography{largevarsR_biblio.bib}

\begin{thebibliography}{33}
\providecommand{\natexlab}[1]{#1}
\providecommand{\url}[1]{\texttt{#1}}
\expandafter\ifx\csname urlstyle\endcsname\relax
  \providecommand{\doi}[1]{doi: #1}\else
  \providecommand{\doi}{doi: \begingroup \urlstyle{rm}\Url}\fi

\bibitem[Anderson(1951)]{Anderson}
T.~W. Anderson.
\newblock Estimating linear restrictions on regression coefficients for
  multivariate normal distributions.
\newblock \emph{Annals of Mathematical Statistics}, 22\penalty0 (3):\penalty0
  327--351, 1951.

\bibitem[Bejan(2005)]{Bejan}
A.~Bejan.
\newblock Largest eigenvalues and sample covariance matrices. {T}racy-widom and
  {P}ainlev\'{e} ii: computational aspects and realization in s-plus with
  applications.
\newblock \emph{Preprint:
  \url{http://users.stat.umn.edu/~jiang040/downloadpapers/largesteigen/largesteigen.pdf}},
  2005.

\bibitem[Bertelli et~al.(2022)Bertelli, Vacca, and Zoia]{bertelli2022bootstrap}
S.~Bertelli, G.~Vacca, and M.~Zoia.
\newblock Bootstrap cointegration tests in ardl models.
\newblock \emph{Economic Modelling}, 116:\penalty0 105987, 2022.

\bibitem[Bornemann(2010)]{bornemann2009numerical}
F.~Bornemann.
\newblock On the numerical evaluation of distributions in random matrix theory:
  a review.
\newblock \emph{Markov Processes Relat. Fields}, 16:\penalty0 803--866, 2010.
\newblock arXiv:0904.1581.

\bibitem[Brockwell and Davis(1991)]{brockwell1991time}
P.~J. Brockwell and R.~A. Davis.
\newblock \emph{Time series: theory and methods}.
\newblock Springer science \& business media, 1991.

\bibitem[Bykhovskaya and Gorin(2022)]{bykhovskaya_gorin_1}
A.~Bykhovskaya and V.~Gorin.
\newblock Cointegration in large {VAR}s.
\newblock \emph{The Annals of Statistics}, 50\penalty0 (3):\penalty0
  1593--1617, 2022.

\bibitem[Bykhovskaya and Gorin(2024)]{BG_review}
A.~Bykhovskaya and V.~Gorin.
\newblock Canonical correlation analysis: review.
\newblock \emph{arXiv preprint arXiv:2411.15625}, 2024.
\newblock \doi{10.48550/arXiv.2411.15625}.

\bibitem[Bykhovskaya and Gorin(2025)]{bykhovskaya_gorin_k}
A.~Bykhovskaya and V.~Gorin.
\newblock Asymptotics of cointegration tests for high-dimensional {VAR}($k$).
\newblock \emph{Review of Economics and Statistics}, 2025.

\bibitem[Dieng(2005)]{dieng2005distribution}
M.~Dieng.
\newblock Distribution functions for edge eigenvalues in orthogonal and
  symplectic ensembles: Painlev{\'e} representations.
\newblock \emph{International Mathematics Research Notices}, 2005\penalty0
  (37):\penalty0 2263--2287, 2005.

\bibitem[Dumitriu and Edelman(2002)]{dumitriu_edelman}
I.~Dumitriu and A.~Edelman.
\newblock Matrix models for beta ensembles.
\newblock \emph{Journal of Mathematical Physics}, 43\penalty0 (11):\penalty0
  5830--5847, 2002.

\bibitem[Edelman and Persson(2005)]{edelman2005numerical}
A.~Edelman and P.-O. Persson.
\newblock Numerical methods for eigenvalue distributions of random matrices.
\newblock \emph{arXiv preprint math-ph/0501068}, 2005.

\bibitem[Engle and Granger(1987)]{engle_granger1987}
R.~Engle and C.~Granger.
\newblock Co-integration and error correction: representation, estimation, and
  testing.
\newblock \emph{Econometrica}, 55\penalty0 (2):\penalty0 251--276, 1987.

\bibitem[Forrester(1993)]{forrester1993}
P.~J. Forrester.
\newblock The spectrum edge of random matrix ensembles.
\newblock \emph{Nuclear Physics B}, 402\penalty0 (3):\penalty0 709--728, 1993.

\bibitem[Gonzalo and Pitarakis(1999)]{gonzalo_pitarakis}
J.~Gonzalo and J.~Y. Pitarakis.
\newblock Dimensionality effect in cointegration analysis.
\newblock In \emph{Cointegration, Causality, and Forecasting. A Festschrift in
  Honour of Clive WJ Granger}, chapter~9, pages 212--229. Oxford University
  Press, Oxford, 1999.

\bibitem[Granger(1981)]{granger1981}
C.~Granger.
\newblock Some properties of time series data and their use in econometric
  model specification.
\newblock \emph{Journal of Econometrics}, 16\penalty0 (1):\penalty0 121--130,
  1981.

\bibitem[Ho and S{\o}rensen(1996)]{ho_sorensen1996}
M.~Ho and B.~E. S{\o}rensen.
\newblock Finding cointegration rank in high dimensional systems using the
  {J}ohansen test: an illustration using data based {M}onte {C}arlo
  simulations.
\newblock \emph{The Review of Economics and Statistics}, 78\penalty0
  (4):\penalty0 726--732, 1996.

\bibitem[Johansen(1988)]{johansen1988}
S.~Johansen.
\newblock Statistical analysis of cointegrating vectors.
\newblock \emph{Journal of Economic Dynamics and Control}, 12\penalty0
  (2--3):\penalty0 231--254, 1988.

\bibitem[Johansen(1991)]{johansen1991}
S.~Johansen.
\newblock Estimation and hypothesis testing of cointegration vectors in
  {G}aussian vector autoregressive models.
\newblock \emph{Econometrica}, 59:\penalty0 1551--1580, 1991.

\bibitem[Johansen(1995)]{johansen_book}
S.~Johansen.
\newblock \emph{Likelihood-based inference in cointegrated vector
  autoregressive models}.
\newblock Oxford University Press, 1995.

\bibitem[Johnstone and Ma(2012)]{johnstone2012fast}
I.~M. Johnstone and Z.~Ma.
\newblock Fast approach to the {Tracy}-{Widom} law at the edge of {GOE} and
  {GUE}.
\newblock \emph{The {A}nnals of {A}pplied {P}robability}, 22\penalty0
  (5):\penalty0 1962, 2012.

\bibitem[Johnstone et~al.(2021)Johnstone, Klochkov, Onatski, and
  Pavlyshyn]{johnstone2021spin}
I.~M. Johnstone, Y.~Klochkov, A.~Onatski, and D.~Pavlyshyn.
\newblock Spin glass to paramagnetic transition in spherical
  {S}herrington-{K}irkpatrick model with ferromagnetic interaction.
\newblock \emph{arXiv preprint arXiv:2104.07629}, 2021.

\bibitem[Johnstone et~al.(2022)Johnstone, Ma, Perry, and
  Shahram]{rmstatpackage}
I.~M. Johnstone, Z.~Ma, P.~O. Perry, and M.~Shahram.
\newblock \emph{RMTstat: Distributions, Statistics and Tests derived from
  Random Matrix Theory}, 2022.
\newblock R package version 0.3.1.

\bibitem[L{\"u}tkepohl et~al.(2004)L{\"u}tkepohl, Saikkonen, and
  Trenkler]{lutkepohl2004testing}
H.~L{\"u}tkepohl, P.~Saikkonen, and C.~Trenkler.
\newblock Testing for the cointegrating rank of a var process with level shift
  at unknown time.
\newblock \emph{Econometrica}, 72\penalty0 (2):\penalty0 647--662, 2004.

\bibitem[Maddala and Kim(1998)]{maddala}
G.~S. Maddala and I.-M. Kim.
\newblock \emph{Unit Roots, Cointegration, and Structural Change}.
\newblock Cambridge University Press, 1998.

\bibitem[Natsiopoulos and Tzeremes(2023)]{ardlpackage}
K.~Natsiopoulos and N.~Tzeremes.
\newblock \emph{ARDL: ARDL, ECM and Bounds-Test for Cointegration}, 2023.
\newblock URL \url{https://CRAN.R-project.org/package=ARDL}.
\newblock R package version 0.2.4.

\bibitem[Onatski and Wang(2018)]{onatski_ecta}
A.~Onatski and C.~Wang.
\newblock Alternative asymptotics for cointegration tests in large vars.
\newblock \emph{Econometrica}, 86\penalty0 (4):\penalty0 1465--1478, 2018.

\bibitem[Onatski and Wang(2019)]{onatski_joe}
A.~Onatski and C.~Wang.
\newblock Extreme canonical correlations and high-dimensional cointegration
  analysis.
\newblock \emph{Journal of Econometrics}, 2019.

\bibitem[Pesaran et~al.(2001)Pesaran, Shin, and Smith]{pesaran2001bounds}
M.~H. Pesaran, Y.~Shin, and R.~J. Smith.
\newblock Bounds testing approaches to the analysis of level relationships.
\newblock \emph{Journal of applied econometrics}, 16\penalty0 (3):\penalty0
  289--326, 2001.

\bibitem[Pfaff(2008)]{urcapackage}
B.~Pfaff.
\newblock \emph{Analysis of Integrated and Cointegrated Time Series with R}.
\newblock Springer, New York, second edition, 2008.
\newblock URL \url{https://www.pfaffikus.de}.
\newblock ISBN 0-387-27960-1.

\bibitem[Phillips and Ouliaris(1990)]{phillips1990asymptotic}
P.~C. Phillips and S.~Ouliaris.
\newblock Asymptotic properties of residual based tests for cointegration.
\newblock \emph{Econometrica: journal of the Econometric Society}, pages
  165--193, 1990.

\bibitem[Tracy and Widom(1996)]{tw1996}
C.~A. Tracy and H.~Widom.
\newblock On orthogonal and symplectic matrix ensembles.
\newblock \emph{Communications in Mathematical Physics}, 177\penalty0
  (3):\penalty0 727--754, 1996.

\bibitem[Trogdon and Zhang(2024)]{trogdon2024computing}
T.~Trogdon and Y.~Zhang.
\newblock Computing the tracy-widom distribution for arbitrary beta.
\newblock \emph{SIGMA. Symmetry, Integrability and Geometry: Methods and
  Applications}, 20:\penalty0 005, 2024.

\bibitem[Vacca and Bertelli(2024)]{bootctpackage}
G.~Vacca and S.~Bertelli.
\newblock \emph{bootCT: Bootstrapping the ARDL Tests for Cointegration}, 2024.
\newblock URL \url{https://CRAN.R-project.org/package=bootCT}.
\newblock R package version 2.1.0.

\end{thebibliography}

\end{document}